\documentclass[10pt,showpacs,preprintnumbers,amsmath,amssymb,prd,nofootinbib,superscriptaddress]{revtex4-2}
\usepackage{dcolumn} 
\usepackage{bm} 
\usepackage{ifpdf} 
\usepackage{hyperref} 
\usepackage{xcolor,color,graphicx,graphics,physics} 
\usepackage[spanish,english]{babel} 
\usepackage[latin1]{inputenc} 
\usepackage[OT1]{fontenc} 
\usepackage{latexsym,amssymb,amsmath,amsfonts, slashed,cancel, simpler-wick} 
\usepackage{makeidx} 
\usepackage{epsfig,subfigure} 
\usepackage{natbib} 
\usepackage{epstopdf} 
\usepackage{mathrsfs} 
\usepackage{hyperref} 
\hypersetup{colorlinks=true, linkcolor=blue, citecolor=blue, urlcolor=blue} 
\usepackage{enumerate} 
\usepackage{tikz} 
\usepackage{feynmp-auto} 
\usepackage{tikz-feynman} 
\tikzfeynmanset{compat=1.1.0} 
\usepackage{fixmath} \everymath{\displaystyle} 
\usepackage{graphicx} 
\usepackage[T1]{fontenc} 
\usepackage{amsmath} 
\usepackage{amssymb} 
\usepackage{graphicx} 
\usepackage{xcolor} 
\newcommand{\bea}{\begin{eqnarray}} 
\newcommand{\eea}{\end{eqnarray}}

\newcommand{\orcid}[1]{\href{https://orcid.org/#1}{\includegraphics[width=10pt]{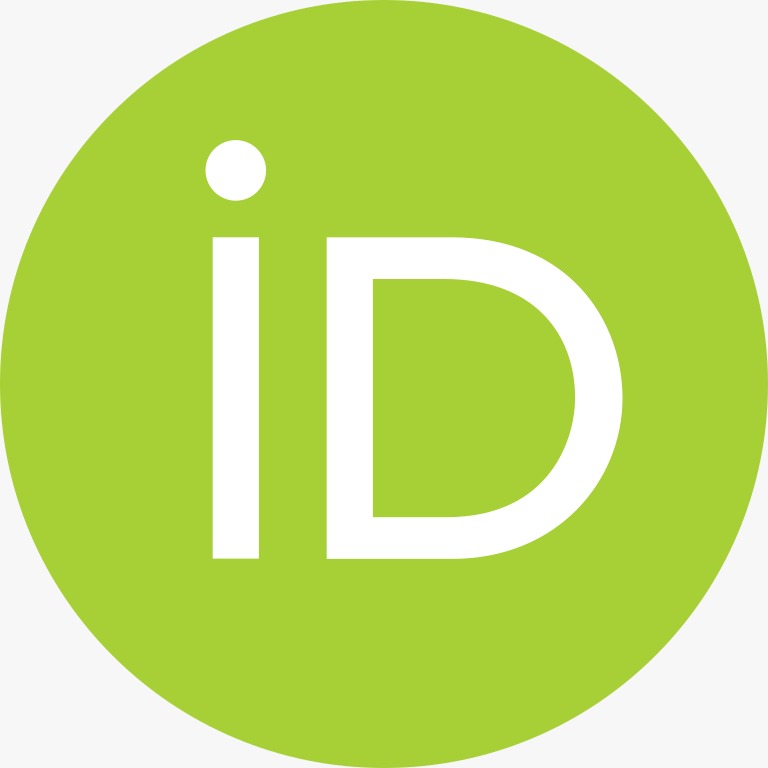}}}

\begin{document}
\baselineskip=18pt

\title{Are There Closed Timelike Curves in $f(R,\mathcal{L}_m,\Phi,g^{\mu\nu}\nabla_\mu \Phi \nabla_\nu \Phi)$-Gravity?}

\author{Faizuddin Ahmed\orcid{0000-0003-2196-9622}}
\email{faizuddinahmed15@gmail.com}
\affiliation{Department of Physics, The Assam Royal Global University, Guwahati, 781035, Assam, India}  

\author{A. F. Santos \orcid{0000-0002-2505-5273}}
\email{alesandroferreira@fisica.ufmt.br (Corresponding author)}
\affiliation{Programa de P\'{o}s-Gradua\c{c}\~{a}o em F\'{\i}sica, Instituto de F\'{\i}sica,\\ 
Universidade Federal de Mato Grosso, Cuiab\'{a}, Brasil}

\begin{abstract}
A modified gravitational model whose action is given by an arbitrary function of the Ricci scalar, the matter Lagrangian density, a scalar field, and its kinetic term is investigated as an extension of the gravitational sector including an additional dynamical degree of freedom. Within this framework, the causal structure of rotating cosmological solutions is analyzed by considering a cylindrically symmetric Pertov-type N space-times and an axially symmetric Petrov type-III with a cosmological constant as background geometries used as theoretical probes of the model consistency. In both cases, pure radiation as matter sources are examined, including a scalar-field configurations. We demonstrate that, although the considered space-times are exact solutions to the field equations of general relativity with a matter source, they are inconsistent within the modified gravity theory considered here.\\

\end{abstract}

\maketitle

\section{Introduction}

Several observational results indicate that general relativity is not a complete theory of gravitation. In particular, it does not provide a satisfactory explanation for the late-time accelerated expansion of the universe, the nature of dark matter, nor does it possess a consistent quantum formulation. These issues motivate the search for alternative theories of gravity or modifications of general relativity. Among the various proposals of modified gravity are $f(R)$ gravity \cite{defelice2010fR}, $f(R, L_m)$ gravity \cite{FRLm}, $f(R,Q)$ gravity \cite{FRQ}, and $f(R,T)$ gravity \cite{FRT}, among others. For detailed reviews on modified theories of gravity, see Refs.~\cite{lima2004alternative,clifton2012modified}.

In this work, we investigate a modified theory of gravity motivated by the possibility that the gravitational interaction may involve nontrivial couplings between geometry, matter, and additional scalar degrees of freedom. Such extensions of general relativity have attracted considerable attention as promising candidates for addressing some of the main open problems in gravitation and cosmology, including the late-time accelerated expansion of the universe and possible deviations from Einstein gravity at high-energy scales. In this context, we consider a generalized gravitational action described by the functional form $f(R,\mathcal{L}_m,\Phi,g^{\mu\nu}\nabla_\mu\Phi\nabla_\nu\Phi)$ \cite{harko2018extensions,Harko_paper,Harko_2}, in which the scalar field is treated as an intrinsic dynamical degree of freedom of the gravitational sector. This framework provides a unified description of the interplay between curvature, matter, and scalar fields, while allowing for a broad class of gravitational dynamics without imposing a specific cosmological origin for the scalar field.

An important aspect in the investigation of modified theories of gravity is whether exact solutions known from general relativity remain valid within the extended gravitational framework. In particular, special attention has been devoted to solutions that admit Closed Timelike Curves (CTCs), since such geometries allow the existence of trajectories returning to the observer's own past, leading to violations of causality. The study of causality violation in modified gravity theories is especially relevant because modifications in the gravitational dynamics may alter the conditions under which CTCs can arise, potentially excluding or enlarging the class of noncausal solutions admitted in general relativity. Indeed, Einstein's theory is known to admit several exact solutions with nontrivial causal structure, among them the well-known G\"odel spacetime \cite{godel1949} (see also, Ref. \cite{Raychaudhuri1955,Reboucas1983}) and its generalizations \cite{Banerji1957,Gurses2005}. In the context of $f(R,\mathcal{L}_m,\Phi,g^{\mu\nu}\nabla_\mu\Phi\nabla_\nu\Phi)$ gravity, G\"odel and G\"odel-type solutions were recently investigated in Ref.~\cite{Evangelista2026}.

In the present work, we employ a cylindrically symmetric Petrov type-N and an axially symmetric type-III space-times with a negative cosmological constant \cite{Ahmed2018,Ahmed2016,deSouza2023,deSouza2024} as background geometries to investigate whether the causality violations present in general relativity persist in the modified gravitational theory under consideration. In the context of general relativity, these solutions are supported by a pure radiation field satisfying the energy conditions together with a negative cosmological constant. A remarkable feature of these geometries is the existence of closed timelike curves in timelike regions of the spacetime, leading to a violation of the causality condition. Moreover, such curves arise dynamically from an initially causal spacelike hypersurface at a finite instant $t=t_0>0$. Motivated by these properties, we analyze the field equations of the $f(R,\mathcal{L}_m,\Phi,g^{\mu\nu}\nabla_\mu\Phi\nabla_\nu\Phi)$ theory by considering both pure radiation and scalar field configurations as possible matter sources.

This paper is organized as follows. In Sec.~\ref{II}, we introduce the action describing the gravitational model and derive the corresponding field equations of $f(R,\mathcal{L}_m,\Phi,g^{\mu\nu}\nabla_\mu\Phi\nabla_\nu\Phi)$ gravity. In Sec.~\ref{III}, we investigate whether the Petrov type-N AdS spacetime, expressed in cylindrical coordinates, constitutes a consistent solution of the modified gravitational theory. In Sec.~\ref{IV}, a similar analysis is carried out for the axially symmetric Petrov type-III AdS spacetime. In both cases, we examine the consistency of the resulting system of field equations and discuss whether the theory admits the existence of Closed Timelike Curves (CTCs). Finally, in Sec.~\ref{V}, we summarize the main results and present our concluding remarks.

\section{Field equations of $f(R,\mathcal{L}_m,\Phi,X)$ gravity}\label{II}

In this section, we present the main equations governing the $f(R,\mathcal{L}_m,\Phi,X)$ gravitational theory. This framework extends the $f(R,\mathcal{L}_m)$ models discussed in Ref.~\cite{FRLm} by incorporating a scalar field as an additional dynamical component of the gravitational sector. Consequently, the matter content is described not only by the standard matter Lagrangian density but also by the presence of a scalar degree of freedom coupled to the geometry.

The action associated with the theory is given by \cite{Evangelista2026}
\begin{eqnarray}
 S= \frac{1}{2 \kappa} \int d^4x\,\sqrt{-g}\left[ f(R,\mathcal{L}_m,\Phi,X)-2\Lambda \right],
 \label{aa1}
\end{eqnarray}
where $g$ denotes the determinant of the metric tensor $g_{\mu\nu}$, $R$ is the Ricci scalar, $\mathcal{L}_m$ represents the matter Lagrangian density, and $\Lambda$ is the cosmological constant. The quantity
\[
X \equiv g^{\mu\nu}\nabla_\mu\Phi\nabla_\nu\Phi
\]
corresponds to the kinetic term of the scalar field $\Phi$. Throughout this work, the function $f(R,\mathcal{L}_m,\Phi,X)$ is assumed to be analytic in all of its arguments, which guarantees the mathematical consistency of the model and permits expansions around a given background spacetime.

The gravitational field equations are obtained by varying the action with respect to the metric tensor, yielding
\begin{align}
f_R R_{\mu\nu}+\left(g_{\mu\nu}\Box-\nabla_\mu\nabla_\nu\right)f_R-\frac{1}{2}\left(f-f_{\mathcal{L}_m}\mathcal{L}_m\right)g_{\mu\nu}
+f_{(\nabla\Phi)^2}\nabla_\mu\Phi\nabla_\nu\Phi
-\frac{1}{2}f_{\mathcal{L}_m} T_{\mu\nu}
+\Lambda g_{\mu\nu}=0,
\label{aa8}
\end{align}
where $f_Y\equiv \partial f/\partial Y$, with
$Y=R,\mathcal{L}_m,\Phi,(\nabla\Phi)^2$, and
$(\nabla\Phi)^2=g^{\mu\nu}\nabla_\mu\Phi\nabla_\nu\Phi$.
The operator
$\Box=g^{\mu\nu}\nabla_\mu\nabla_\nu$
denotes the covariant d'Alembertian. These equations generalize Einstein's field equations by allowing nonminimal couplings among curvature, matter, and the scalar kinetic sector.

The equation governing the scalar field is obtained from the variation of the action with respect to $\Phi$, leading to
\begin{align}
\Box_{(\nabla\Phi)^2}\Phi
=\frac{1}{2}f_\Phi,
\label{11}
\end{align}
where the generalized differential operator is defined as
\begin{align}
\Box_{(\nabla\Phi)^2}
=\frac{1}{\sqrt{-g}}\partial_\mu
\left(f_{(\nabla\Phi)^2}\sqrt{-g}\,g^{\mu\nu}\partial_\nu\right).
\end{align}
This relation can be interpreted as a generalized Klein--Gordon equation associated with the modified gravitational dynamics.

In order to investigate explicit solutions, we consider the particular functional form
\begin{equation}
f = R+\mathcal{L}_m + \frac{\lambda}{2}
g^{\mu\nu}\nabla_\mu\Phi\nabla_\nu\Phi,
\label{aa13}
\end{equation}
where $\lambda$ is a constant coupling parameter. Moreover, we restrict the analysis to the case of a vanishing scalar potential, which simplifies the dynamical equations and allows the effects of the scalar kinetic contribution to be analyzed separately.

Substituting Eq.~(\ref{aa13}) into Eq.~(\ref{aa8}), the field equations become
\begin{equation}
R_{\mu\nu}
-\frac{1}{2}R\,g_{\mu\nu}
+\Lambda g_{\mu\nu}
=
\kappa T_{\mu\nu}
+\frac{\lambda}{4}
g_{\mu\nu}
g^{\sigma\tau}
(\nabla_\sigma\Phi)(\nabla_\tau\Phi)
-\frac{\lambda}{2}
(\nabla_\mu\Phi)(\nabla_\nu\Phi).
\label{aa15}
\end{equation}

For the choice given by Eq.~(\ref{aa13}), the scalar field equation reduces to
\begin{equation}
\frac{1}{\sqrt{-g}}
\partial_\mu
\left(
\frac{\lambda}{2}
\sqrt{-g}\,
g^{\mu\nu}\partial_\nu\Phi
\right)=0.
\label{aa16}
\end{equation}
Using the identities
\begin{align}
\Gamma^\alpha_{\alpha\mu}
=
\frac{\partial_\mu\sqrt{-g}}{\sqrt{-g}},
\qquad
\partial_\mu g^{\mu\nu}
=
-\Gamma^{\mu}_{\mu\alpha}g^{\alpha\nu}
-\Gamma^{\nu}_{\mu\alpha}g^{\mu\alpha},
\label{aa17}
\end{align}
Eq.~(\ref{aa16}) can be rewritten as
\begin{align}
g^{\mu\nu}
\left(
\partial_\mu\partial_\nu\Phi
-\Gamma^\alpha_{\mu\nu}\partial_\alpha\Phi
\right)=0,
\label{aa18}
\end{align}
which is equivalent, up to the constant factor $\lambda$, to the standard covariant wave equation $\Box\Phi=0$. Therefore, within this particular model, the scalar field obeys the usual massless Klein--Gordon equation in curved spacetime.

In the next sections, the modified field equations (\ref{aa15}) are applied to axially symmetric Petrov type-III and type-N metrics in the presence of a cosmological constant.

\section{Petrov type-N Cylindrical Symmetry AdS Space-time}\label{III}

The line-element that describes Petrov type-N AdS space-time in the cylindrical coordinates $(t, r, \phi, z)$ is given by the following expression (assuming $c = 1$ and $8\pi G = 1$) \cite{Ahmed2018}:
\begin{equation}
ds^2 = g_{rr}\,dr^2 + 2 g_{t\phi}\,dt\,d\phi + g_{\phi\phi}\,d\phi^2 + 2 g_{z\phi}\,dz\,d\phi + g_{zz}\,dz^2,\label{bb1}
\end{equation}
with different components of the metric tensor $g_{\mu\nu}$ given by
\begin{align}
g_{t\phi} &= -\frac{1}{2} \cosh t \sinh^2(\alpha r),\nonumber\\
g_{rr} &= \coth^2(\alpha r), \nonumber\\
g_{\phi\phi} &= -\sinh t \sinh^2(\alpha r),\nonumber\\
g_{zz} &= \sinh^2(\alpha r), \nonumber\\
g_{z\phi} &= \beta z \sinh^2(\alpha r),\label{bb2}
\end{align}
where $\alpha > 0$, $\beta > 0$ are arbitrary positive constants.

To convert this metric (\ref{bb1}) with (\ref{bb2}) into a standard form, we perform transformations as follows
\begin{equation}
t \to \sinh^{-1}(\tau), \quad r \to \frac{1}{\alpha}\sinh^{-1}(\alpha \varrho),\label{bb3}
\end{equation}
into the metric (\ref{bb1}), we obtain the following line-element in the chart $(\tau, \varrho, \phi, z)$ given by
\begin{equation}
ds^2 = \frac{d\varrho^2}{\alpha^2 \varrho^2} + \alpha^2 \varrho^2 \left( - d\tau\, d\phi - \tau\, d\phi^2 + 2 \beta z\, dz\, d\phi + dz^2 \right).\label{bb4}
\end{equation}

The metric tesnor and its contrvariant form are given by
\begin{equation}
    g_{\mu\nu} =
\begin{pmatrix}
0 & 0 & -\frac{1}{2}\alpha^{2}\varrho^{2} & 0 \\
0 & \frac{1}{\alpha^{2}\varrho^{2}} & 0 & 0 \\
-\frac{1}{2}\alpha^{2}\varrho^{2} & 0 & -\alpha^{2}\varrho^{2}\tau & \alpha^{2}\varrho^{2}\beta z \\
0 & 0 & \alpha^{2}\varrho^{2}\beta z & \alpha^{2}\varrho^{2}
\end{pmatrix},\quad g^{\mu\nu} =
\begin{pmatrix}
\frac{4\tau + \beta^{2}z^{2}}{\alpha^{2}\varrho^{2}} & 0 & -\frac{2}{\alpha^{2}\varrho^{2}} & \frac{\beta z}{\alpha^{2}\varrho^{2}} \\
0 & \alpha^{2}\varrho^{2} & 0 & 0 \\
-\frac{2}{\alpha^{2}\varrho^{2}} & 0 & 0 & 0 \\
\frac{\beta z}{\alpha^{2}\varrho^{2}} & 0 & 0 & \frac{1}{\alpha^{2}\varrho^{2}}
\end{pmatrix}.
\end{equation}

The Ricci scalar for the above metric (\ref{bb4}) is given by
\begin{equation}
    R=g^{\mu\nu} R_{\mu}=-12 \alpha^2.\label{bb5}
\end{equation}
Moreover, the Ricci tensor is given by
\begin{equation}
R_{\mu\nu} =\begin{pmatrix}
0 & 0 & \dfrac{3 \alpha^4 \varrho^2}{2} & 0 \\
0 & -\dfrac{3}{\varrho^2} & 0 & 0 \\
\dfrac{3 \alpha^4 \varrho^2}{2} & 0 & \beta + 3 \tau \alpha^4 \varrho^2 & -3 \alpha^4 \beta z \varrho^2 \\
0 & 0 & -3 \alpha^4 \beta z \varrho^2 & -3 \alpha^4 \varrho^2
\end{pmatrix}.\label{bb6}
\end{equation}

The non-zero components of the Christoffel symbols are given by
\begin{align}
&\Gamma^{\tau}{}_{\tau\varrho}
=\Gamma^{\tau}{}_{\varrho\tau}
=\frac{1}{\varrho},\quad 
\Gamma^{\tau}{}_{\varrho\phi}
=\Gamma^{\tau}{}_{\phi\varrho}
=\frac{2\tau+\beta^{2}z^{2}}{\varrho},\quad 
\Gamma^{\tau}{}_{\varrho z}
=\Gamma^{\tau}{}_{z\varrho}
=-\frac{\beta z}{\varrho},\nonumber\\
&\Gamma^{\varrho}{}_{\varrho\varrho}
=-\frac{1}{\varrho},\quad 
\Gamma^{\varrho}{}_{\tau\phi}
=\Gamma^{\varrho}{}_{\phi\tau}
=\frac{1}{2}\alpha^{4}\varrho^{3},\quad 
\Gamma^{\varrho}{}_{\phi\phi}
=\alpha^{4}\varrho^{3}\tau,\quad 
\Gamma^{\varrho}{}_{\phi z}
=\Gamma^{\varrho}{}_{z\phi}
=-\alpha^{4}\varrho^{3}\beta z,\quad
\Gamma^{\varrho}{}_{zz}
=-\alpha^{4}\varrho^{3},\nonumber\\
&\Gamma^{\phi}{}_{\varrho\tau}
=\Gamma^{\phi}{}_{\tau\varrho}
=\frac{1}{\varrho},\quad \Gamma^{\phi}{}_{\varrho\phi}
=\Gamma^{\phi}{}_{\phi\varrho}
=\frac{1}{\varrho},\quad \Gamma^{z}{}_{\varrho z}
=\Gamma^{z}{}_{z\varrho}
=\frac{1}{\varrho},\quad \Gamma^{z}{}_{\phi z}
=\Gamma^{z}{}_{z\phi}
=\beta.
\end{align}

To solve the scalar field equation (\ref{aa18}), the ansatz $\Phi=\Phi(\varrho)$ is assumed. Under this assumption, Eq. (\ref{aa18}) reduces to
\begin{equation}
    g^{11} (\Phi'(\varrho))^2-g^{11}\,\Gamma^{1}_{11} (\Phi'(\varrho))=0\Longrightarrow (\Phi'(\varrho))^2+\frac{1}{\varrho}\Gamma^{1}_{11} (\Phi'(\varrho))=0\label{bb7}
\end{equation}
whose solution is given by
\begin{equation}
    \Phi (\varrho)=a \ln\!\varrho.\label{bb8}
\end{equation}

Using this, we find the kinetic term as follows:
\begin{equation}
    g^{\mu\nu} (\nabla_{\mu} \phi) (\nabla_{\nu} \Phi)=(\nabla \Phi)^2=g^{11} (\Phi'(\varrho))^2=\frac{a^2}{\varrho^2}\,g^{11}=a^2 \alpha^2.\label{bb9}
\end{equation}

To solve the modified field equations, we consider the energy-momentum a pure radiation field, whose energy-momentum tensor is given by
\begin{equation}
    T_{\mu\nu}=\rho k_{\mu} k_{\nu},\quad k_{\mu}=(0, 0, 1 ,0),\quad k^{\mu} k_{\mu}=0.\label{cc8}
\end{equation}

Therefore, the field equations (\ref{aa15}) in terms of components are as follows:
\begin{align}
    &\frac{3}{\varrho^2}+\frac{\Lambda}{\alpha^2 \varrho^2}=\frac{\lambda a^2}{4 \varrho^2}-\frac{\lambda a^2}{2},\label{bb10}\\
    &-\frac{3 \alpha^4}{2} \varrho^2-\frac{\Lambda}{2}\alpha^2 \varrho^2=-\frac{\lambda}{2} a^2 \alpha^2 \varrho^2,\label{bb11}\\
    &\beta-3 \alpha^4 \varrho^2 \tau-\Lambda \alpha^2 \varrho^2 \tau=\rho-\frac{\lambda}{4} \alpha^4 \varrho^2 a^2 \tau,\label{bb12}\\
    &3 \alpha^4 \beta \varrho^2 z+\Lambda \alpha^2 \beta \varrho^2 z=\frac{\lambda}{4} \alpha^4 a^2 \beta \varrho^2 z,\label{bb13}\\
    & 3 \alpha^4 \varrho^2+\Lambda \alpha^2 \varrho^2=\frac{\lambda}{4} \alpha^4 \varrho^2a^2.\label{bb14} 
\end{align}

By comparing Eqs.~(\ref{bb11}) and (\ref{bb14}), one finds that the system of equations is inconsistent, since the field equations do not admit compatible solutions for the cosmological constant and the energy density, in contrast to what occurs in general relativity. Consequently, the modified gravity theory considered here does not allow solutions of the form given in Eq.~(\ref{bb1}). Moreover, the absence of consistent solutions indicates that the causality violation commonly associated with this metric, namely the existence of closed timelike curves, does not arise within this theoretical framework.

Alternatively, one may interpret this result as an indication that the modifications introduced in the gravitational sector impose stronger geometrical constraints than those present in general relativity, preventing the emergence of rotating cosmological solutions that violate causality.

\section{Petrov type-III Axially symmetric AdS Space-time}\label{IV}

Here, we consider an axially symmetric metric admitting Closed Timelike Curves (CTCs) in general relativity~\cite{Ahmed2018,Ahmed2016,deSouza2023,deSouza2024}. The line element that describes this spacetime at $(x^0=t,\, x^1=r,\, x^2=\phi,\, x^3=z)$ coordinates is given as
\begin{equation}
ds^2 = \frac{dr^2}{\alpha^2 r^2} + r^2 dz^2 + \left( -2r^2 dt + \frac{\beta z\, dr}{r^2} - t r^2 d\phi \right) d\phi,\label{cc1}
\end{equation}
where $\alpha$ and $\beta$ are non-zero constants, with $\beta > 0$. It should be noted from Eq.~(\ref{cc1}) that this spacetime has a coordinate 
singularity at $r = 0$.

The metric tensor and its contravariant form are given by
\begin{equation}
    g_{\mu\nu} =
\begin{pmatrix}
0 & 0 & -r^{2} & 0 \\
0 & \dfrac{1}{\alpha^{2} r^{2}} & \dfrac{\beta z}{2 r^{2}} & 0 \\
- r^{2} & \dfrac{\beta z}{2 r^{2}} & -r^{2} t & 0 \\
0 & 0 & 0 & r^{2}
\end{pmatrix},\quad g^{\mu\nu} =
\begin{pmatrix}
\frac{\alpha^{2}\beta^{2} z^{2}}{4 r^{6}} + \frac{t}{r^{2}} & \frac{\alpha^{2}\beta z}{2 r^{2}} & -\frac{1}{r^{2}} & 0 \\
\frac{\alpha^{2}\beta z}{2 r^{2}} & \alpha^{2} r^{2} & 0 & 0 \\
-\frac{1}{r^{2}} & 0 & 0 & 0 \\
0 & 0 & 0 & \frac{1}{r^{2}}
\end{pmatrix}.\label{metric-tensor}
\end{equation}

The Ricci tensor for the above space-time is given by ($\mu,\nu=0, \cdots, 3$)
\begin{equation}
R_{\mu\nu} =\begin{pmatrix}
0 & 0 & 3\alpha^{2} r^{2} & 0 \\
0 & -\dfrac{3}{r^{2}} & -\dfrac{3\alpha^{2}\beta z}{2r^{2}} & 0 \\
3\alpha^{2} r^{2} & -\dfrac{3\alpha^{2}\beta z}{2r^{2}} & \dfrac{\alpha^{2}\left(\beta^{2}+24 r^{6} t\right)}{8r^{4}} & 0 \\
0 & 0 & 0 & -3\alpha^{2} r^{2}
\end{pmatrix}.\label{cc2}
\end{equation}
The Ricci scalar is given as
\begin{align} 
R=g^{\mu \nu }R_{\mu \nu }= & {} -12 \alpha ^2.\label{cc3} 
\end{align}

The non-zero components of the Christoffel symbols $\Gamma^{\lambda}_{\mu\nu}$ are as follows:
\begin{align}
\Gamma^{0}{}_{01} &= \frac{1}{r}, \qquad
\Gamma^{0}{}_{02} = \frac{\alpha^{2}\beta z + r}{2r}, \qquad
\Gamma^{0}{}_{11} = \frac{\beta z}{2r^{5}}, \quad 
\Gamma^{0}{}_{12} = -\frac{\alpha^{2}\beta^{2} z^{2}}{4r^{5}}, \qquad
\Gamma^{0}{}_{13} = -\frac{\beta}{4r^{4}}, \nonumber\\[4pt]
\Gamma^{0}{}_{22} &= \frac{\alpha^{2}\beta^{2} z^{2} + 4\alpha^{2}\beta r^{3}tz + 4r^{4}t}{8r^{4}}, \qquad
\Gamma^{0}{}_{23} = \frac{\alpha^{2}\beta^{2} z}{8r^{4}}, \quad 
\Gamma^{0}{}_{33}= -\frac{\alpha^{2}\beta z}{2r}, \qquad
\Gamma^{1}{}_{02} = \alpha^{2} r^{3}, \nonumber\\[4pt]
\Gamma^{1}{}_{12}&= -\frac{\alpha^{2}\beta z}{2r}, \qquad
\Gamma^{1}{}_{22} = \frac{1}{4}\alpha^{2}\left(\beta z + 4r^{3}t\right),\quad 
\Gamma^{1}{}_{23}= \frac{\alpha^{2}\beta}{4}, \qquad
\Gamma^{1}{}_{33} = -\alpha^{2} r^{3}, \qquad
\Gamma^{2}{}_{12} = \frac{1}{r}, \nonumber\\[4pt]
\Gamma^{2}{}_{22} &= -\frac{1}{2}, \qquad
\Gamma^{3}{}_{12} = -\frac{\beta}{4r^{4}}, \qquad
\Gamma^{3}{}_{13} = \frac{1}{r},\quad \Gamma^{1}{}_{11} = -\frac{1}{r}.\label{cc4}
\end{align}

To solve the scalar field equation (\ref{aa18}), the ansatz $\Phi=\Phi(r)$ is assumed. Under this assumption, Eq. (\ref{aa18}) reduces to
\begin{equation}
    g^{11} (\Phi'(r))^2-g^{11}\,\Gamma^{1}_{11} (\Phi'(r))=0\label{cc5}
\end{equation}
whose  solution is given by 
\begin{equation}
    \Phi(r)=b \ln\! r.\label{cc6}
\end{equation}
Using this, we find the kinetic term as follows:
\begin{equation}
    g^{\mu\nu} (\nabla_{\mu} \Phi) (\nabla_{\nu} \Phi)=(\nabla \Phi)^2=g^{11} (\Phi'(r))^2=\frac{b^2}{r^2}\,g^{11}=\alpha^2 b^2.\label{cc7}
\end{equation}

With these, the modified field equations using the energy-momentum tensor (\ref{cc8}) can be expressed as
\begin{align}
&3 \alpha^2 r^2+\Lambda r^2=\frac{\lambda\,b^2}{4}\,\alpha^2 r^2,\label{cc9}\\
&3 \alpha^2 r^2+\Lambda r^2=\frac{\lambda\,b^2}{4}\,\alpha^2 r^2-\frac{\lambda}{2}\,\frac{b^2}{r^2},\label{cc10}\\
&3 \alpha^2 r^2+\Lambda r^2=\frac{\lambda\,b^2}{4} \,\alpha^2 r^2,\label{cc11}\\
&\frac{\alpha^2 \beta^2}{8 r^4}-3 \alpha^2 r^2 t-r^2 t \Lambda=\rho+\frac{\lambda b^2 t}{4} \,\alpha^2 r^2,\label{cc12}\\
&3 \alpha^2 r^2+\Lambda r^2=\frac{\lambda\,b^2}{4}\,\alpha^2 r^2,\label{cc13}
\end{align}

By analyzing Eqs. (\ref{cc10}) and (\ref{cc13}), it is immediately found that the resulting system of equations is inconsistent. This inconsistency arises from the contribution of the scalar field. In particular, when $\lambda\,b^2=0$, the scalar-field kinetic term vanishes and the field equations reduce to those of general relativity. In this limit, the well-known axially symmetric solution is recovered, for which the energy density and the cosmological constant are given by
\begin{equation}
    \Lambda=-3 \alpha^2,\quad \rho=\frac{\alpha^2 \beta^2}{\rho r^4}.\label{cc14}
\end{equation}
Therefore, the inconsistency of the modified field equations indicates that the axially symmetric Petrov-type spacetime admitting Closed Timelike Curves in general relativity is no longer supported within the $f(R,\mathcal{L}_m,\phi,g^{\mu\nu}\nabla_\mu\phi\nabla_\nu\phi)$ framework. In particular, the contribution associated with the scalar-field sector prevents the existence of consistent solutions satisfying the required geometrical and matter conditions. Consequently, for the class of solutions considered here, the modified gravitational dynamics prohibit the occurrence of Closed Timelike Curves, thereby avoiding the corresponding violation of causality present in general relativity.

\section{Conclusions}\label{V}

In this work, we investigated the causal structure of the $f(R,\mathcal{L}_m,\phi,g^{\mu\nu}\nabla_\mu\phi\nabla_\nu\phi)$ theory of gravity, which has been proposed as an extension of general relativity incorporating nonminimal couplings between curvature, matter, and scalar fields. This framework allows for a broad class of gravitational dynamics while providing a unified description of the interplay between geometric and scalar degrees of freedom, without requiring a specific cosmological interpretation for the scalar field.

A central aspect of the analysis was the investigation of whether spacetime geometries admitting Closed Timelike Curves (CTCs) in general relativity remain valid solutions in the modified theory. The existence of CTCs is closely related to the violation of causality and has long been one of the most intriguing features of Einstein gravity. In order to address this issue, we considered a cylindrically symmetric Petrov type-N AdS and an axially symmetric Petrov type-III spacetimes as background geometries, both sourced by pure radiation fields in the presence of a cosmological constant.

Our results indicate that, within the $f(R,\mathcal{L}_m,\phi,g^{\mu\nu}\nabla_\mu\phi\nabla_\nu\phi)$ framework considered here, the corresponding systems of field equations become inconsistent for both the space-times under consideration. In particular, unlike the situation in general relativity, no compatible solutions for the cosmological constant and the radiation energy density can be obtained. Consequently, these Petrov-type geometries do not constitute admissible solutions of the theory. Therefore, for the class of spacetimes analyzed in this work, the modified gravitational dynamics prevent the occurrence of causality violation associated with the existence of Closed Timelike Curves. However, it is worth noting that these space-time geometries under investigation are exact solutions of another modified gravity theory, namely Ricci-inverse gravity, and therefore still violate the causality condition in this framework (see Refs.~\cite{deSouza2023,deSouza2024,Ahmed2024,Santos2024}).

\scriptsize

\section*{Acknowledgments}

F.A. acknowledges the Inter University Centre for Astronomy and Astrophysics (IUCAA), Pune, India for granting visiting associateship. This work by A. F. Santos is partially supported by National Council for Scientific and Technological Development - CNPq project No. 312406/2023-1.

\section*{Data Availability Statement}

No Data associated in the manuscript.

\section*{Conflicts of Interest}

No conflict of interest in this paper.

\bibliographystyle{apsrev4-2}
\bibliography{reference.bib} 

\end{document}